\documentclass[preprint,12pt]{elsarticle}
\usepackage[utf8]{inputenc}
\usepackage{xcolor}
\usepackage{amssymb}
\usepackage{amsmath}
\begin{document}

\begin{frontmatter}

%% Title, authors and addresses

%% use the tnoteref command within \title for footnotes;
%% use the tnotetext command for theassociated footnote;
%% use the fnref command within \author or \affiliation for footnotes;
%% use the fntext command for theassociated footnote;
%% use the corref command within \author for corresponding author footnotes;
%% use the cortext command for theassociated footnote;
%% use the ead command for the email address,
%% and the form \ead[url] for the home page:
%% \title{Title\tnoteref{label1}}
%% \tnotetext[label1]{}
%% \author{Name\corref{cor1}\fnref{label2}}
%% \ead{email address}
%% \ead[url]{home page}
%% \fntext[label2]{}
%% \cortext[cor1]{}
%% \affiliation{organization={},
%%             addressline={},
%%             city={},
%%             postcode={},
%%             state={},
%%             country={}}
%% \fntext[label3]{}

\title{Enhancing hydrogen production in alkaline electrolyzers by using an ultra-fast kinetics surfactant} %% Article title

\author[label1,label2]{D. Fernández-Martínez}
\author[label1]{G. Martín-Zazo}
\author[label1,label2]{A. Rubio-González}
\author[label3]{J. L. Serrano-Marcos}
\author[label1,label2]{J. M. Montanero}
\author[label3]{F. J. Perosanz}
\author[label1,label2]{E. J. Vega\corref{cor1}}

\address[label1]{Depto. de Ingeniería Mecánica, Energética y de los Materiales, Universidad de Extremadura, E-06006 Badajoz, Spain}
\address[label2]{Instituto de Computación Científica Avanzada (ICCAEx), Universidad de Extremadura, E-06006 Badajoz, Spain}
\address[label3]{Unidad de Conversión Electroquímica de Energía-H2 (UCEEH2), Departamento de Energía CIEMAT, Madrid, Spain}
\cortext[cor1]{Corresponding author: E. J. Vega, \texttt{ejvega@unex.es}}

%% Abstract
\begin{abstract}
We study the effect of surfactant adsorption kinetics on water electrolysis performance using Surfynol 465, an ultrafast-kinetics surfactant. High-speed optical diagnostics reveal that Surfynol 465 reduces bubble residence time by an order of magnitude. Compared to the slower surfactant Triton X-100 and the surfactant-free baseline, it also prevents the growth of large bubbles ($>200\ \mu$m). We validated these results on an anion-exchange membrane electrolyzer (AEMEL) test bench. Adding Surfynol 465 to the $1\text{ M KOH}$ electrolyte decreases cell overpotential by 140--150 mV. It nearly triples the current density (from 0.13 to 0.37 A/cm$^2$ at 2 V) and increases average power by 40\% during a week-long test. Furthermore, downstream corrosion analyses reveal a strong alloy-dependent response. The surfactant reduces degradation in stainless steel and brass but accelerates corrosion in carbon steel.
\end{abstract}

%% Keywords
\begin{keyword}

water electrolysis \sep surfactants \sep hydrogen bubble dynamic \sep hydrogen production rate

\end{keyword}

\end{frontmatter}

%% Add \usepackage{lineno} before \begin{document} and uncomment 
%% following line to enable line numbers
%% \linenumbers

%% main text

\section{Introduction}
\label{sec1}
% Bubble adhesion in electrolysis
Water electrolysis powered by renewable energy sources represents a major pathway toward clean hydrogen production. Among the available technologies, alkaline water electrolysis (AWE) and anion exchange membrane water electrolysis (AEMWE or AEMEL, anion exchange membrane electrolyzer) are highly attractive due to their cost-effectiveness and robustness \citep{T24,SGLXJZQ25,SABLKHRH24, RNSMKAY25}. However, their operating efficiency remains limited by transport phenomena in the electrodes \citep{WWYLZTZYA26}. During operation, the rapid generation of hydrogen gas produces a dense population of bubbles that adhere to the catalytic surface. This bubble accumulation isolates the electrode from the electrolyte, causing fluctuations in the reaction potential, increasing local ohmic resistance, and reducing the electrochemically active surface area \citep{MIF12, JJWYY26}. Consequently, the development of effective methods to eliminate or reduce these adverse effects and accelerate detachment is widely recognized as a crucial step to minimize these energy losses \citep{SBSADA26,ZXZWXZZTCLS25}.

% Available strategies to enhance bubble detachment
Strategies to regulate bubble detachment can be broadly classified into active and passive methods \citep{JJWYY26,ZXZWXZZTCLS25}. Active regulation involves introducing external energy sources, such as applying ultrasonic fields to induce acoustic flow \citep{ANM24} or external magnetic fields to trigger magnetic convection \citep{BG25,YESU08}. Although effective, these active strategies require additional equipment, thereby increasing the cost of the water electrolysis system \citep{ZXZWXZZTCLS25,KCL24}. In contrast, passive regulation methods do not require external energy. These primarily focus on altering the electrode morphology to minimize the contact area between the electrode and the bubble \citep{SBSADA26}. However, manufacturing these complex geometries on an industrial scale remains a challenge. Alternatively, adding surfactants to the electrolyte solution represents a highly practical and cost-effective passive strategy to enhance bubble detachment \citep{LNYLZMJ24,BC18,WXLYGZZ23}.

% Physics of bubble detachment in electrolysis: the role of surfactants
An adhered bubble becomes unstable and detaches when the forces promoting release, such as buoyancy, inertial force, and pressure force, overcome the forces preventing detachment. This resistance is primarily driven by the surface tension force and the Marangoni force resulting from local thermal or solutal gradients. Mechanically, the liquid-gas surface tension $\sigma$ directly dictates both the capillary anchoring and the strength of the Marangoni stress. By accumulating at the interface, surfactants lower $\sigma$, weakening these resistive forces and reducing the bubble departure diameter. Surfactants are conventionally evaluated based on their equilibrium surface tension $\sigma_{eq}$, assuming a fully developed interfacial layer. Nevertheless, in many dynamic processes, the actual speed of surfactant adsorption dictates the effective surface tension during the event, making the evolution of this reduction more critical than the final equilibrium value \citep{QSBBBS20,VSQCC24}.

\begin{figure*}[tbp]
\begin{center}
\resizebox{0.8\textwidth}{!}{\includegraphics{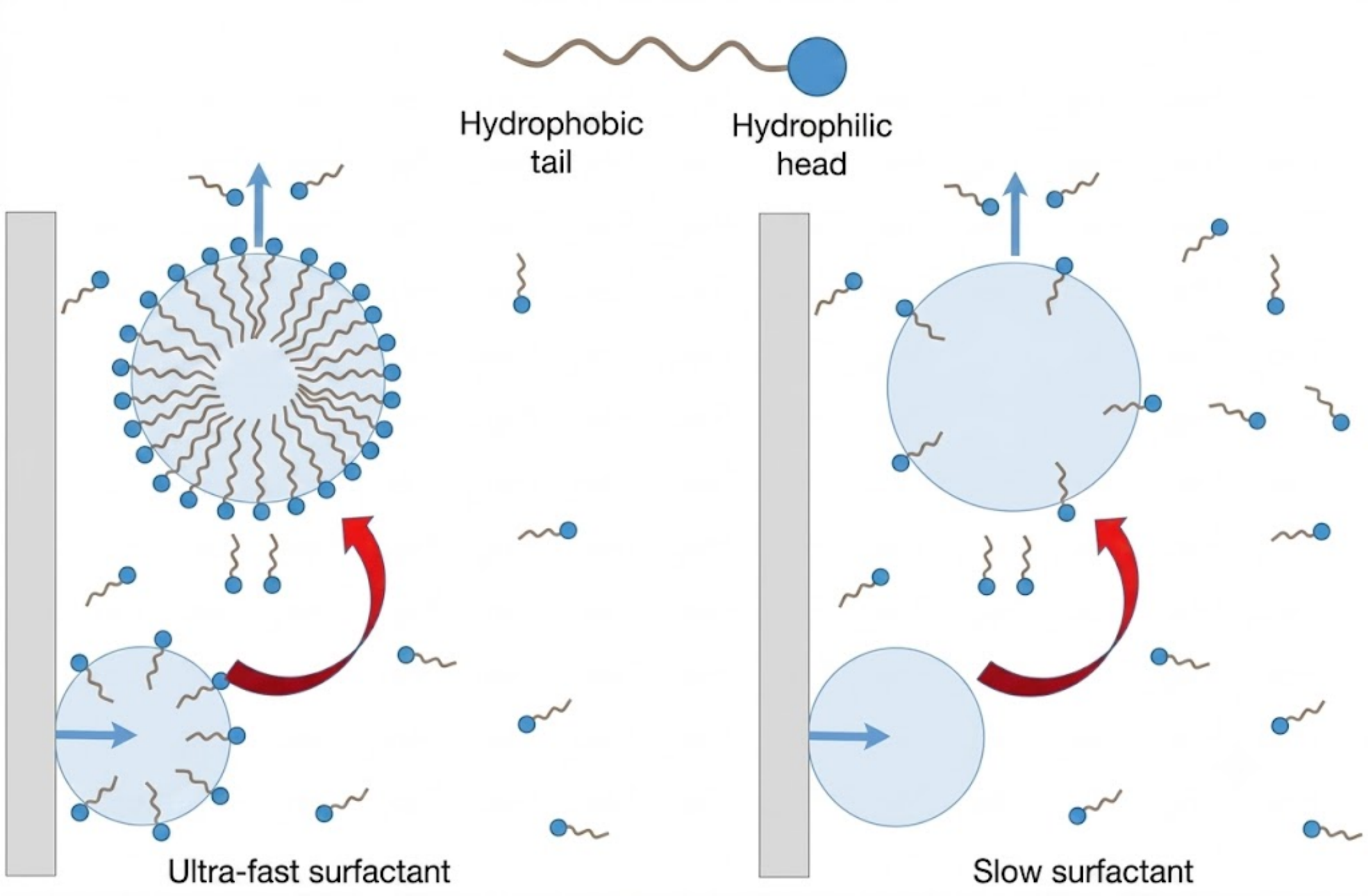}}   
\end{center}
\caption{Sketch showing surfactant adsorption during hydrogen bubble nucleation.}
\label{surf}
\end{figure*}

% Time-scale of the problem and DST
Driven by rapid growth and coalescence, an electrolytic bubble nucleates, expands, and detaches on characteristic timescales spanning from a few milliseconds down to the sub-millisecond regime \citep{ZXZWXZZTCLS25,SSE13,DSLDFO26}. When a bubble grows this rapidly, a new liquid-gas interface is generated at a rate that may outpace the transport and adsorption of surfactant molecules from the bulk solution \citep{MS20,M24} (Fig.\ \ref{surf}). As a result, the local surface tension at the expanding interface changes dynamically over time. This transient behavior is described by the dynamic surface tension (DST) \citep{ED00,KQBMWW26}. If the surfactant molecules diffuse and adsorb too slowly, the interface could remain depleted of surfactant during the bubble's short lifetime \citep{VSQCC24}. In such a scenario, the surfactant might fail to assist bubble detachment, regardless of how low its surface tension is at equilibrium.

% Open question
Despite its potential relevance, the role of surfactant adsorption kinetics and dynamic surface tension in water electrolysis remains largely unexplored. To the best of our knowledge, no prior studies have investigated how the timescale of surfactant adsorption couples with the rapid expansion of gas bubbles under electrolytic conditions. Whether ``ultrafast'' surfactants \citep{FVMGF26} can noticeably improve bubble detachment in water electrolysis compared to slower alternatives remains an open question. %Although both types may achieve equivalent equilibrium surface tension and interfacial coverage, the impact of the adsorption kinetics under electrolytic conditions has yet to be determined.

% What we have done
This work aims to address this question by establishing a link between surfactant kinetics and macroscopic electrolyzer performance. To achieve this, we evaluate the performance of two non-ionic surfactants: Surfynol 465 as an ultrafast candidate and Triton X-100 as a standard, slower alternative. First, a simple laboratory-scale cell is used to conduct high-speed optical diagnostics, enabling us to compare bubble detachment dynamics in the pure electrolyte with those in surfactant-laden solutions. Our experiments reveal that Surfynol 465 substantially modifies the electrolysis process, yielding clear improvements in bubble residence time and hydrogen production compared to Triton X-100 and baseline solutions. Second, the performance of Surfynol 465 is validated under more realistic operating conditions in a larger-scale AEMEL electrolyzer test bench. These validation tests confirm a significant reduction in cell overpotential, improved stability, increased energetic efficiency, and reduced corrosion kinetics compared to the baseline electrolyte.

% Structure
The manuscript is organized as follows. The experimental methodology is described in Secs.\ \ref{sec2} and \ref{sec3}. Section \ref{sec4} shows the high-speed optical diagnostics using a simple laboratory-scale cell, while the results of the more realistic AEMEL electrolyzer are shown in Sec.\ \ref{sec5}. The paper concludes with remarks in Sec.\ \ref{sec6}.

\section{Materials and methods for high-speed optical diagnostics}
\label{sec2}

\subsection{Experimental setup, surfactants and experimental procedure}

% Experimental setup
A simplified AEMEL-type (AWE) electrolyzer model was implemented using a custom-made methacrylate cell allowing optical access to the process. The cell has internal dimensions of 40×40×40 mm and contains approximately 50 ml of solution. The electrodes were approximately 40 mm high and 3 mm wide, with an immersion length of about 20 mm, corresponding to an immersion surface area of 60 mm$^2$. They were made of sintered stainless-steel fiber felt, a porous material characterized by high mechanical, thermal, and corrosion resistance, and a three-dimensional reticulated structure.

\begin{figure*}[tbp]
\begin{center}
\resizebox{1\textwidth}{!}{\includegraphics{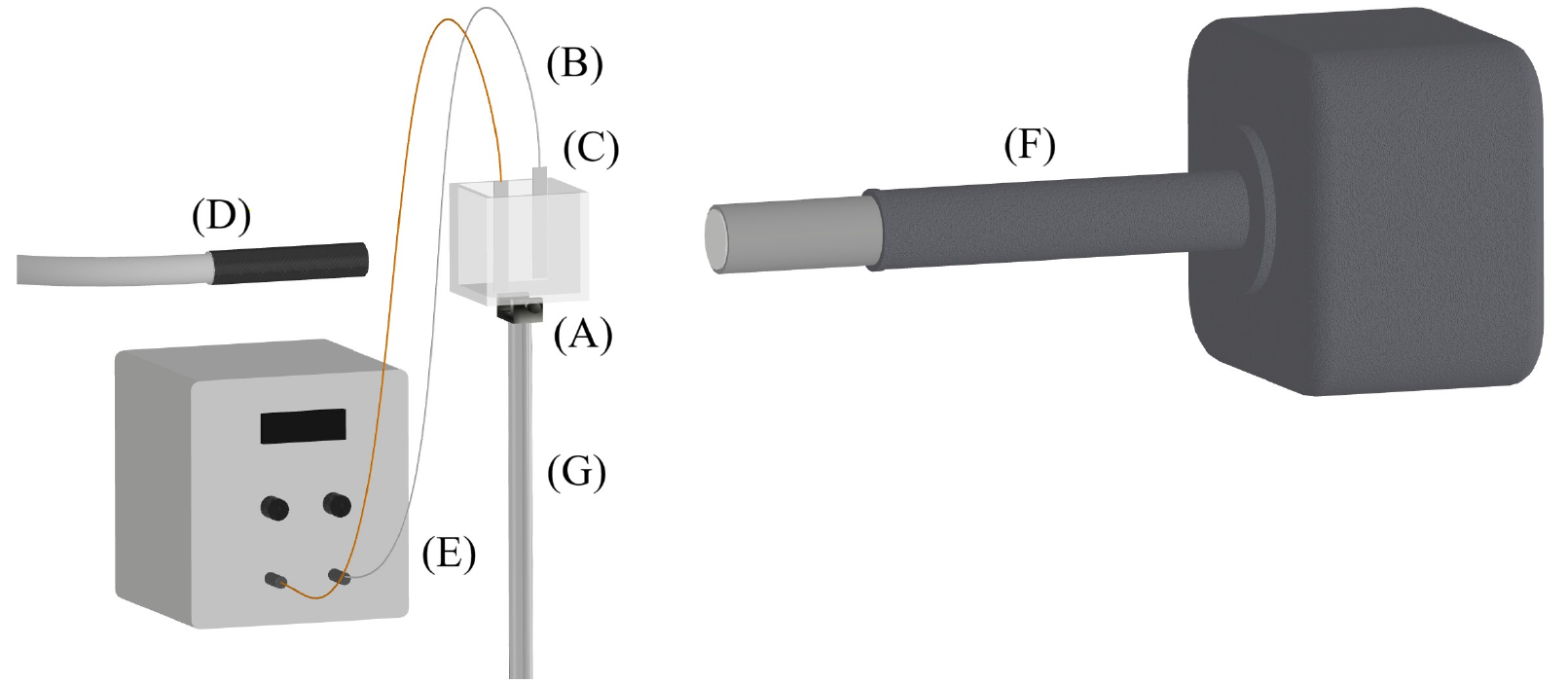}}   
\end{center}
\caption{Experimental setup: cell (A), electrical wiring (B), electrodes (C), light source (D), power supply (E), high-speed camera (F), adjustable support (G).}
\label{setup}
\end{figure*}

Figure \ref{setup} shows the experimental apparatus, consisting of the experimental cell (A), electrical wiring (B), electrodes (C), a light source (D), a power supply (E), a high-speed camera (F), and a height-adjustable support (G). Electrolysis was driven by applying a potential difference between the electrodes using a laboratory power supply. Electrical connections were made using cables with crocodile clips and standard terminals. Two multimeters were used to monitor the applied voltage and the electrical current. Particular care was taken to maintain the cleanliness of all components to ensure reproducible experimental conditions.

Bubble dynamics were recorded at 125 frames per second with an exposure time of 12.5 $\mu$s using a high-speed CMOS camera ({\sc Photron, FASTCAM Mini UX50}). The camera was equipped with an optical system consisting of a 1× zoom objective ({\sc OPTEM HR}) combined with an additional lens system ({\sc OPTEM 70 XL}) providing 3× magnification, resulting in a spatial resolution of 2.315 $\mu$m per pixel. A dedicated light source provided uniform illumination of the observation region. All components were mounted on an optical table equipped with a pneumatic vibration isolation system ({\sc Thorlabs}). Experiments were conducted at an ambient temperature of approximately 20\, $^{\circ}$C.

% Liquids
The working fluid consisted of ultrapure water (W) (Type III) and potassium hydroxide (KOH) at a concentration of 1 M. The addition of KOH increases the electrical conductivity of the solution W+KOH, while the use of ultrapure water minimizes the presence of impurities. For the concentrations considered, the density and viscosity remain close to those of pure water. Ultrapure water was supplied by a {\sc Direct-Q®3} purification system.

\begin{figure*}[tbp]
\begin{center}
\resizebox{0.5\textwidth}{!}{\includegraphics{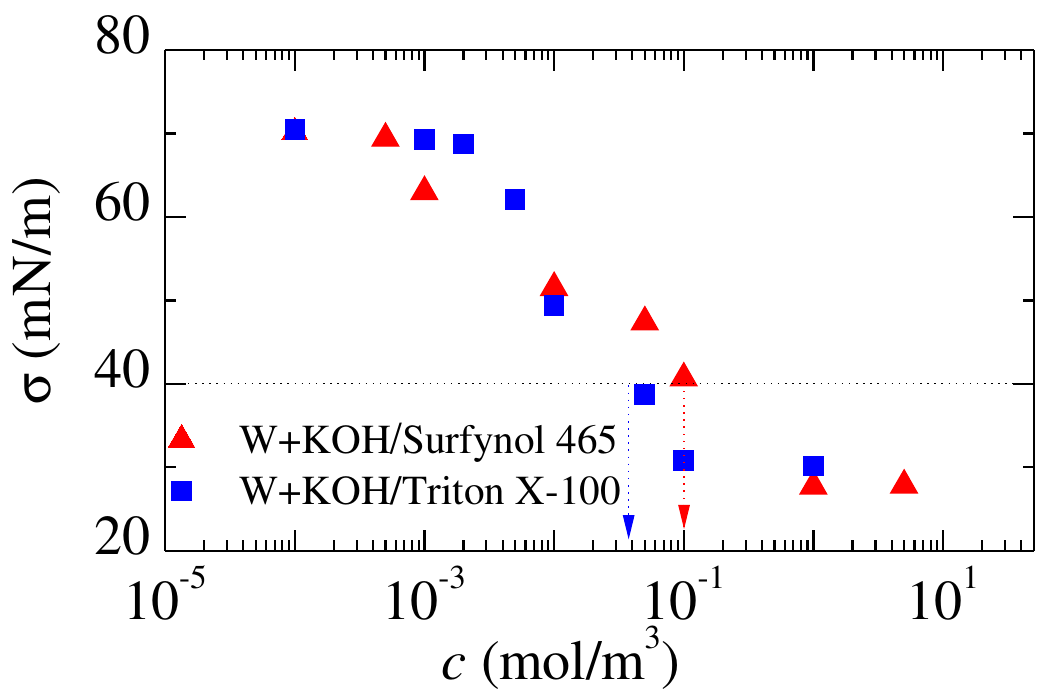}}   
\end{center}
\caption{Surface tension $\sigma$ as a function of the surfactant volumetric concentration \textit{c} for W+KOH/Triton X-100 and W+KOH/Surfynol 465. The arrows indicate the selected concentrations.}
\label{st}
\end{figure*}

Either Triton X-100 or Surfynol 465 was added to the base solution. The surface tension $\sigma$ of both surfactants dissolved in the aqueous electrolytic solution was characterized as a function of the surfactant concentration $c$ (Fig.\ref{st}). From these curves, the critical micelle concentration $c_{cmc}$ was determined to be 0.08 mol/m$^3$ for Triton X-100 and 0.21 mol/m$^3$ for Surfynol 465. A preliminary analysis showed that the effect of Surfynol 465 was maximized for $c=0.1$ mol/m$^3$. We selected this concentration for the rest of our analysis. The Triton X-100 concentration $c=0.038$ mol/m$^3$ was chosen to ensure the same equilibrium surface tension of approximately 40 mN/m. Remarkably, this yields an identical concentration ratio $c/c_{cmc} \approx 0.48$. 

Surfynol 465 rapidly adsorbs at the gas–liquid interface and efficiently reduces surface tension under dynamic conditions \citep{RRMH26}. Its presence is expected to affect bubble size, coalescence, and adhesion to the electrode surface, thereby facilitating bubble detachment. In contrast, the adsorption characteristic time of Triton X-100 is significantly longer than that of Surfynol 465. The differences observed in the presence of the two surfactants can be attributed to their adsorption rates.

To determine the appropriate experimental procedure, the polarization curve was measured over three consecutive voltage ramps (Fig.\ \ref{VI}), with a 1-minute rest at 0.9 V between ramps. On each ramp, the voltage was increased from 1.3 to 2.7 V in 0.2 V steps every 2 minutes. The results of ramps 2 and 3 practically overlapped over almost the entire potential range (Fig.\ \ref{VI}). This indicates that once ramp 1 was completed, the electrode interface had reached an adequate state of stabilization and maturation, in which heterogeneity and material-history effects were eliminated, thereby ensuring maximum reproducibility in subsequent long-term tests.

\textbf{\begin{figure*}[tbp]
\begin{center}
\resizebox{0.5\textwidth}{!}{\includegraphics{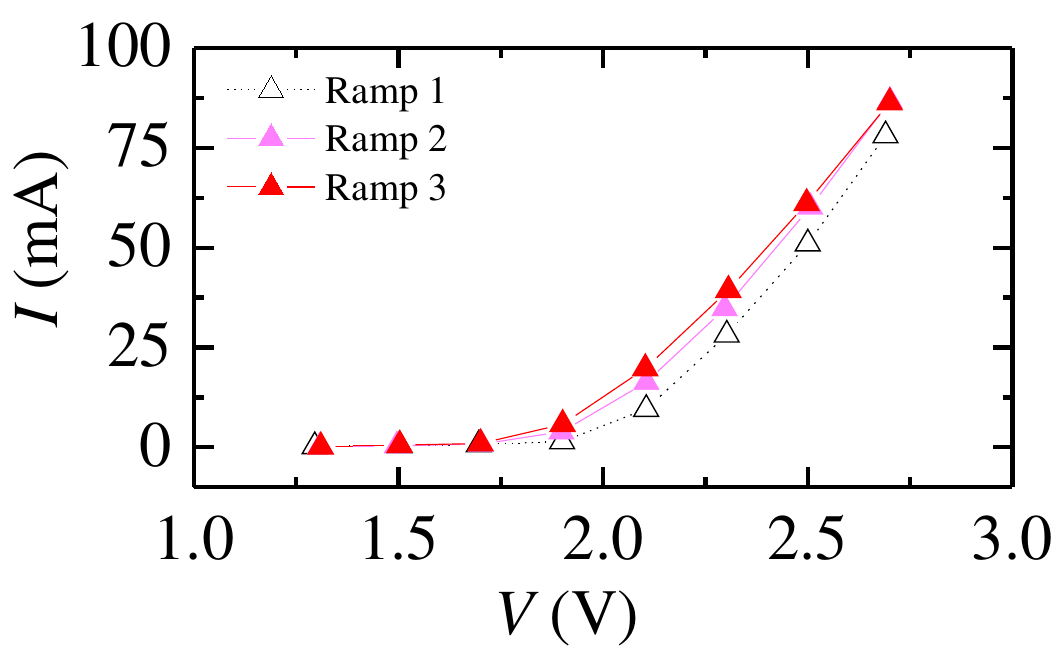}}   
\end{center}
\caption{Polarization curves $I(V)$ for the solution with Surfynol 465.}
\label{VI}
\end{figure*}}

In the activation-loss region ($V\lesssim 2$ V), $I$ is very low due to the energy required to overcome the metal's initial chemical inertia and activate its catalytic sites. The rate of hydrogen generation is minimal, and the bubbles do not obstruct the surface. For $V\gtrsim 2$ V, the system transitioned to a quasi-linear behavior (the ohmic-resistance-loss region). In this regime, the initial chemical barriers have already been overcome, and the current is primarily controlled by the resistance of both the liquid electrolyte and the gas generated as it passes. 

Based on the results above, we followed the following experimental procedure. In each experiment, the solution was first prepared, and new electrodes were installed. The cell was left at rest for 15 minutes to wet the stainless steel felt (hydration), followed by a thermal stabilization phase at 1.9 V for 10 minutes and an activation potential ramp from 1.3 V to 2.7 V (ramp 1). The voltage was then adjusted to a permanent operating value of 2.2 V, marking the start of the experimental run ($t=0$). Six consecutive videos of bubble dynamics at the lower margins of the cathode (three of each face/side) were recorded at specific instants over the 48-hour duration of the experiment. The videos were analyzed using in-house MATLAB code to determine the bubble-size distribution, the fraction of the electrode surface free of bubbles, and the hydrogen production rate, as explained in the next section. 

\subsection{Image processing and evaluation metrics}

% Intro
As noted earlier, $H_2$ bubble dynamics were analyzed via high-speed optical imaging. Dense bubble curtains, coalescence, and overlapping pose significant challenges to tracking individual bubble trajectories. To address this, an automated image-processing framework was developed in MATLAB to characterize the spatial gas occupancy and its collective transport dynamics. It should be noted that the images analyzed represent two-dimensional projections of an inherently three-dimensional bubbly flow. Consequently, the extracted metrics are not absolute physical quantities. Instead, they serve as relative indices for comparative analysis between different electrolytes.

% Stagnant and ascending gas fraction
The primary objective of the algorithm is to decouple the stagnant gas fraction from the ascending one. This physical separation is achieved by evaluating the temporal persistence of the gas phase close to the electrode. The algorithm applies a moving time window centered on each frame. Pixels exhibiting high local occupancy fraction within this window are classified as persistent, stagnant gas. 

\begin{figure*}[tbp]
\begin{center}
\resizebox{1\textwidth}{!}{\includegraphics{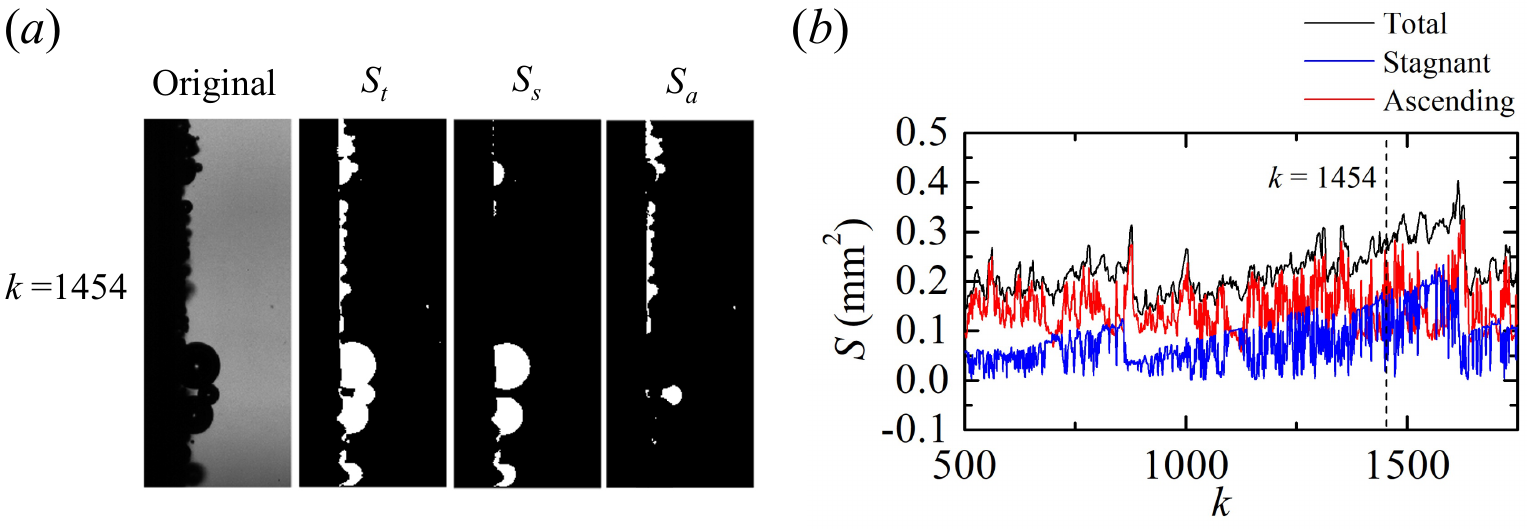}}   
\end{center}
\caption{(a) Binary masks identifying the total $S_t$, stagnant $S_s$, and ascending $S_a$ gas surfaces near the electrode wall at frame $k = 1454$. (b) Surface area $S$ as a function of the frame number $k$. The vertical dashed line indicates the instant corresponding to the masks shown in (a).}
\label{post}
\end{figure*}

The decoupling of the stagnant gas fraction from the ascending one is obtained by the generated binary masks (Fig.\ \ref{post}a), which isolate the total ($S_t$), stagnant ($S_s$), and ascending ($S_a$) gas surfaces near the electrode wall. The algorithm accuracy is demonstrated by tracking these projected surface areas across each frame $k$ of the sequence (Fig. \ref{post}b). While the total gas area series $S_t(k)$ reflects the continuous accumulation and intermittent detachment of large structures, the ascending gas series $S_a(k)$ exhibits remarkably stationary behavior.

Once the areas $S_t(k)$ and $S_a(k)$ have been calculated, we compute the ratio between the mean ascending gas area and the mean total gas area (free-wall percentage $Z$) as
\begin{equation}
Z = \frac{\overline{S}_a}{\overline{S}_t} \times 100
\end{equation}
where $\overline{S}_a=N^{-1}\sum_{k=1}^{N} S_a(k)$ and $\overline{S}_t=N^{-1} \sum_{k=1}^{N} S_t(k)$, and $N$ is the number of frames considered.

% Velocity
Now, we consider the ascending gas fraction to determine the ``bubble curtain's vertical velocity''. To do this, the image is divided into $M$ horizontal windows, 192 pixels in height. For each window $i$, the two-dimensional gas distribution is reduced to a one-dimensional vertical profile by horizontally averaging the binary values (1 for gas, 0 for liquid) of the ascending gas mask. As in Particle Image Velocimetry (PIV), the $i$th window vertical velocity $v_i$ is determined by cross-correlating these profiles across different time steps \citep{RWWK07}. To ensure statistical robustness, the local velocity is filtered using the median over multiple time intervals. Finally, the bubble curtain's vertical velocity $v_b$ is calculated as
\begin{equation}
v_b=\frac{1}{M}\sum_{i=1}^{M} v_i,
\end{equation}

% Metrics
The gas evacuation efficiency is evaluated through the hydrogen production rate index
    \begin{equation}
        i_{Q} = \overline{S_a} \cdot v_b,
    \end{equation}
and the mean bubble residence time 
    \begin{equation}
       t_{r} = \frac{\overline{S}_s}{i_Q},
    \end{equation}
where $\overline{S}_s=N^{-1}\sum_{k=1}^{N} S_s(k)$. We evaluated the energetic efficiency of the electrochemical process by the index 
    \begin{equation}
        i_E = \frac{I \cdot V}{i_Q}.
    \end{equation}
Finally, we also developed a routine to isolate individual bubble contours and calculate the bubble size distribution across the experiments.

\section{Materials and methods for the AEMEL electrolyzer experiments}
\label{sec3}

\begin{figure*}[tbp]
\begin{center}
\resizebox{1\textwidth}{!}{\includegraphics{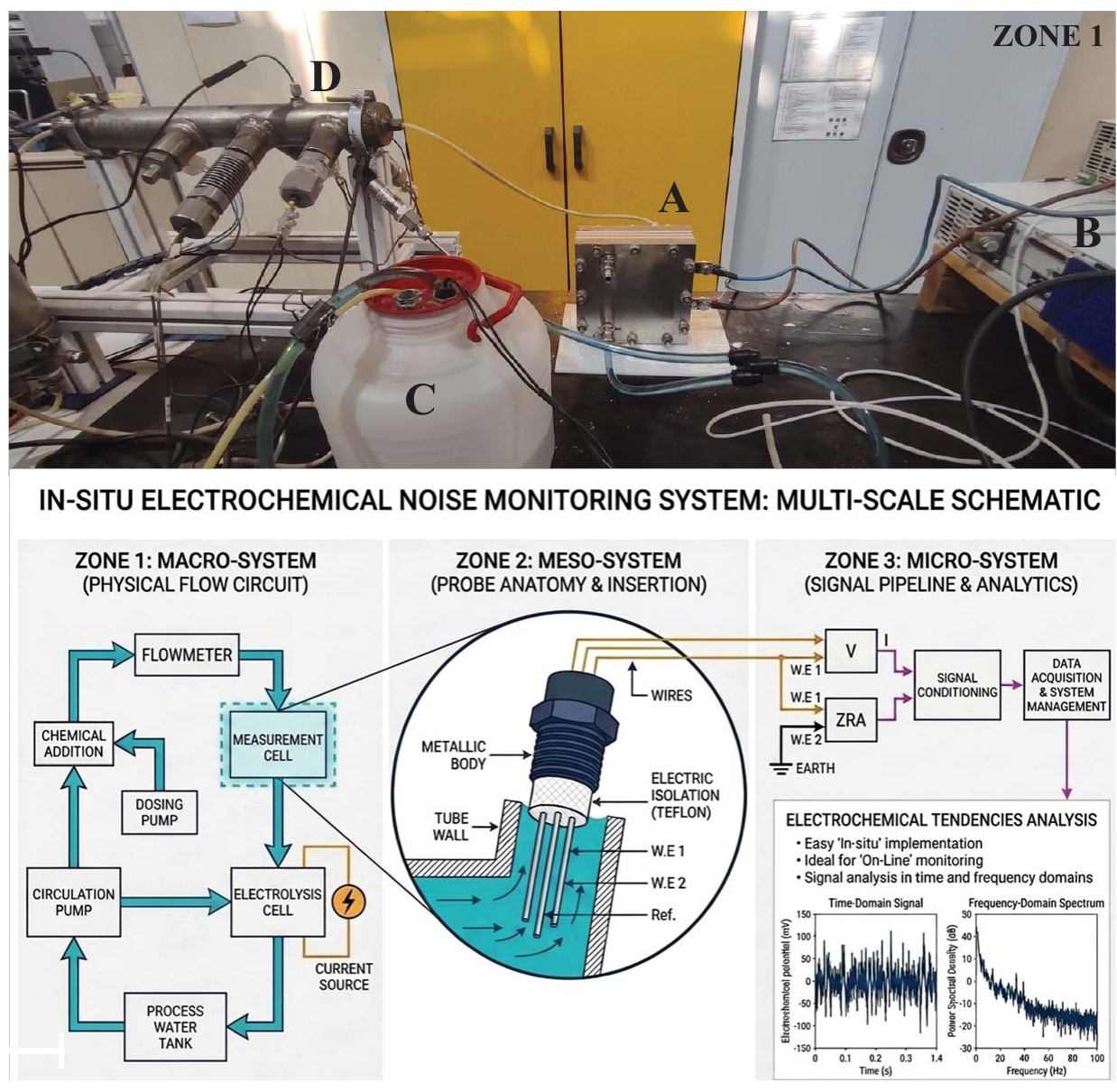}}   
\end{center}
\caption{(Above) Experimental setup for the AEMEL electrolyzer: single cell (A), power source (B), tank (C), probes (D). (Below) Scheme of the electrochemical corrosion testing bench with probes of several alloys for online corrosion/water contamination diagnostic.}
\label{setupAEMEL}
\end{figure*}

Electrochemical evaluation was performed using a single cell Anion Exchange Membrane Electrolysis (AEMEL) system (Fig.\ \ref{setupAEMEL}-above) configured in a zero gap architecture with a dry cathode concept and circular geometry on both the anode and cathode sides, providing a geometric active area of 63.6 cm$^2$. The cell (A) incorporated a commercial anion exchange membrane ({\sc Fumasep\textsuperscript{\textregistered}}), compressed symmetrically between two porous transport layers: a stainless steel mesh (around 0.3 mm in thickness) acting as the gas diffusion layer (GDL), and a nickel foam (3 mm in thickness) serving as the porous transport layer (PTL). No catalyst layers were employed in any of the experiments. Stainless steel meshes were used directly in contact with the membrane, representing a fully catalyst-free configuration.

The electrolyte consisted of 1 M KOH and was continuously recirculated using a chemically resistant positive-displacement pump to ensure uniform hydration and temperature distribution across the membrane–electrode assembly. The power source (B) operated in potentiostatic mode following a predefined activation and stabilization protocol: (i) 24 hours of electrolyte circulation for membrane hydration and homogenization; (ii) 7 days of baseline operation under fixed voltage control; (iii) and sequential potentiostatic sweeps ($I$-$V$ curves) performed at the beginning, midpoint, and end of the test campaign, using a ramp rate of 0.1 V/min to track the evolution of cell performance. We conducted experiments with the tank (C) filled with W+KOH and W+KOH/Surfynol 465. 

In addition, material degradation under alkaline conditions was quantified using downstream electrochemical corrosion probes (D) installed on the anodic outlet line (Fig.\ \ref{setupAEMEL}-below). Each probe consisted of a metallic rod representative of industrially relevant alloys:

\begin{itemize}
    \item[-] {Cu based alloy: Muntz metal (Cu–Zn–Sn), typical in brazing and industrial fittings.}
    \item[-] {Fe based alloy: A106 carbon steel ($Fe > 98\%, C \approx 0.3\%, Mn \approx 0.3-1\%$), representative of high pressure piping.}
    \item[-] {Stainless steel: 316L TIG welding rod (Fe–Cr–Ni–Mo), compatible with alkaline structural components.}
\end{itemize}

Each through-wall probe housed three electrodes made of the same alloy: a reference electrode electrically connected to the stainless steel cell body, a potential measurement electrode, and a working electrode. This configuration establishes a galvanic couple between the alloy and the stainless steel cell, enabling simultaneous measurement of the free corrosion potential $V_{corr}$ and galvanic corrosion current $I_{corr}$ using a high impedance acquisition system equipped with Zero Resistance Ammeters (ZRA) (Fig.\ \ref{setupAEMEL}-below). Signals were recorded continuously throughout the 7-day experiment. The time-integrated corrosion charge $Q$, derived from $I_{corr}$, was used to estimate the kinetics of material loss.

Please note that, although a precise conversion of charge into mass loss would require defining an electrochemical equivalent for each alloy, the galvanic corrosion current measured by the ZRA is directly proportional to the rate of material dissolution. Consequently, $Q$ provides a robust comparative indicator of corrosion kinetics across the different alloys, allowing the identification of trends and relative susceptibility even without applying a full faradaic quantification.

\section{Results. High-speed optical diagnostics}
\label{sec4}

\begin{figure*}[tbp]
\begin{center}
\resizebox{0.5\textwidth}{!}{\includegraphics{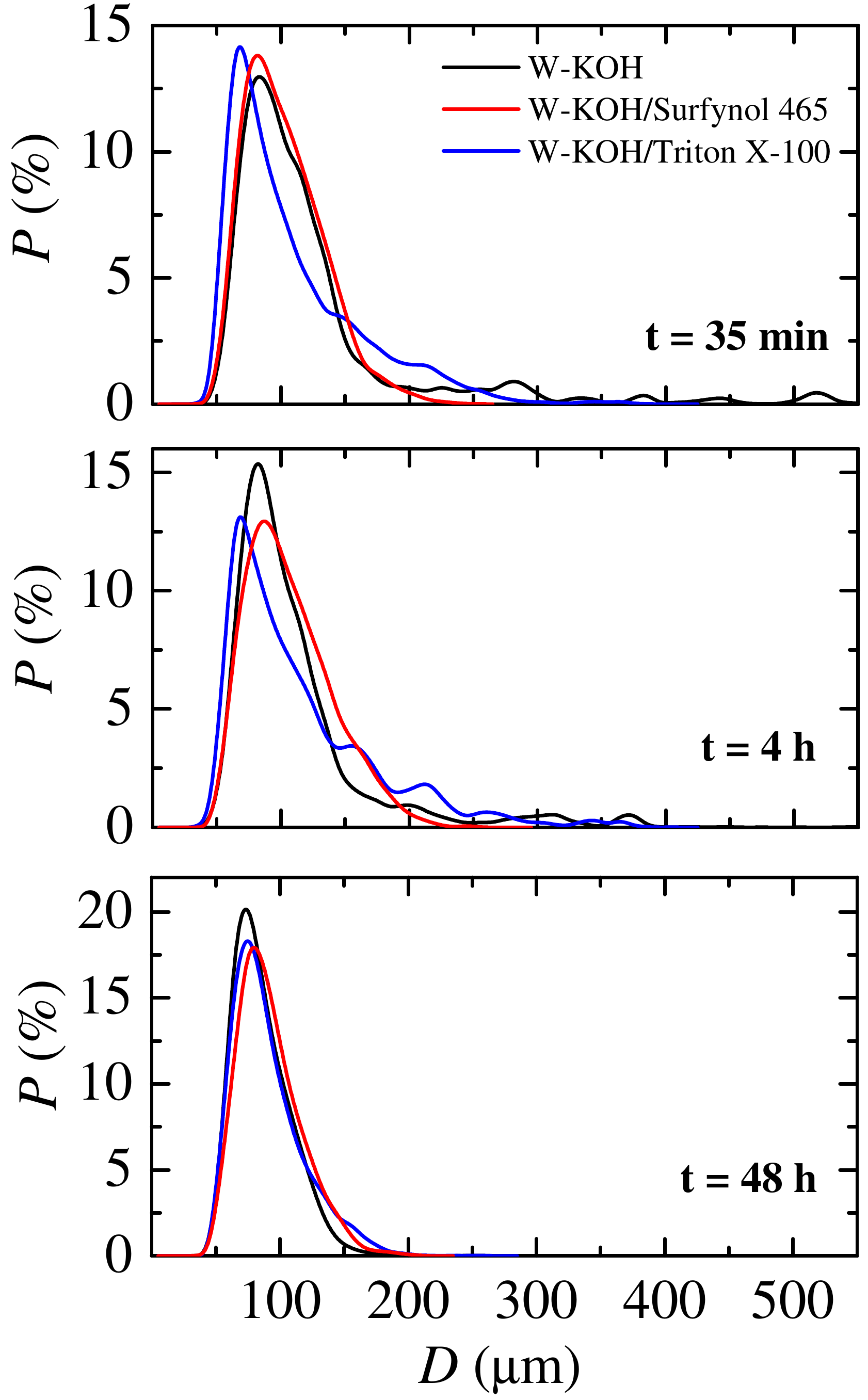}}   
\end{center}
\caption{Probability distribution $P$($D$) of the bubble diameter $D$ for W+KOH, W+KOH/Triton X-100, and W+KOH/Surfynol 465.}
\label{P}
\end{figure*}

\begin{figure*}[tbp]
\begin{center}
\resizebox{0.6\textwidth}{!}{\includegraphics{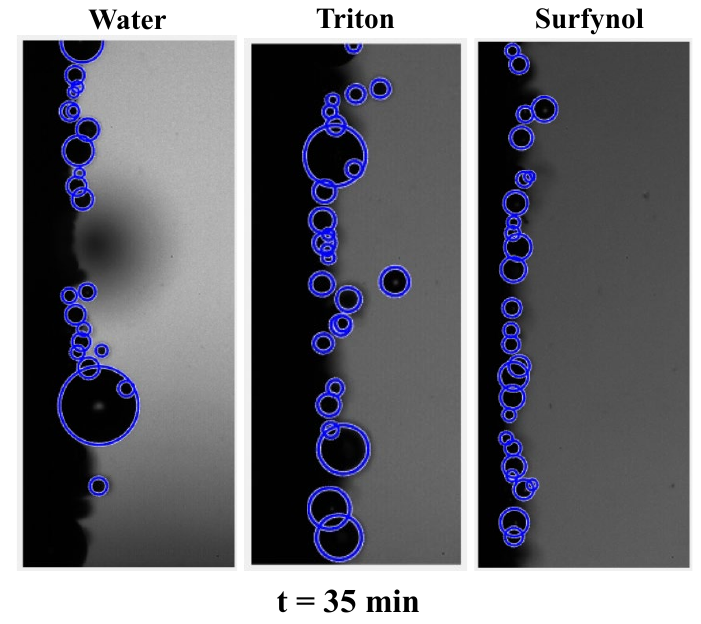}}   
\end{center}
\caption{Snapshot showing the bubble sizes.}
\label{Diameter}
\end{figure*}

This section presents the main results of our high-speed optical diagnostics using a simple laboratory-scale cell described in Sec.\ \ref{sec2}. Hydrogen bubble size is expected to affect electrolyzer efficiency, since large bubbles on the electrode mask the electrode's active area, increasing ohmic losses. We determined the bubble size distribution histograms at $t=35$ min, 4 h, and 48 h for W+KOH, W+KOH/Triton X-100, and W+KOH/Surfynol 465 (Fig.\ \ref{P}). As observed, the surfactant has no significant effect on the overall bubble size distribution. In the three cases, the mean diameter is of the order of 100 $\mu$m with a noticeable degree of polydispersity. However, adding Surfynol 465 prevents the growth of bubbles larger than approximately 200 $\mu$m at $t=35$ min and 4 h. This can also be observed in Fig.\ \ref{Diameter}, which shows the interfacial control exerted by Surfynol 465 during the first phase of the process. It should be noted that the large bubbles produced in the absence of Surfynol 465 constitute a significant fraction of the gas volume.

\begin{figure*}[tbp]
\begin{center}
\resizebox{1\textwidth}{!}{\includegraphics{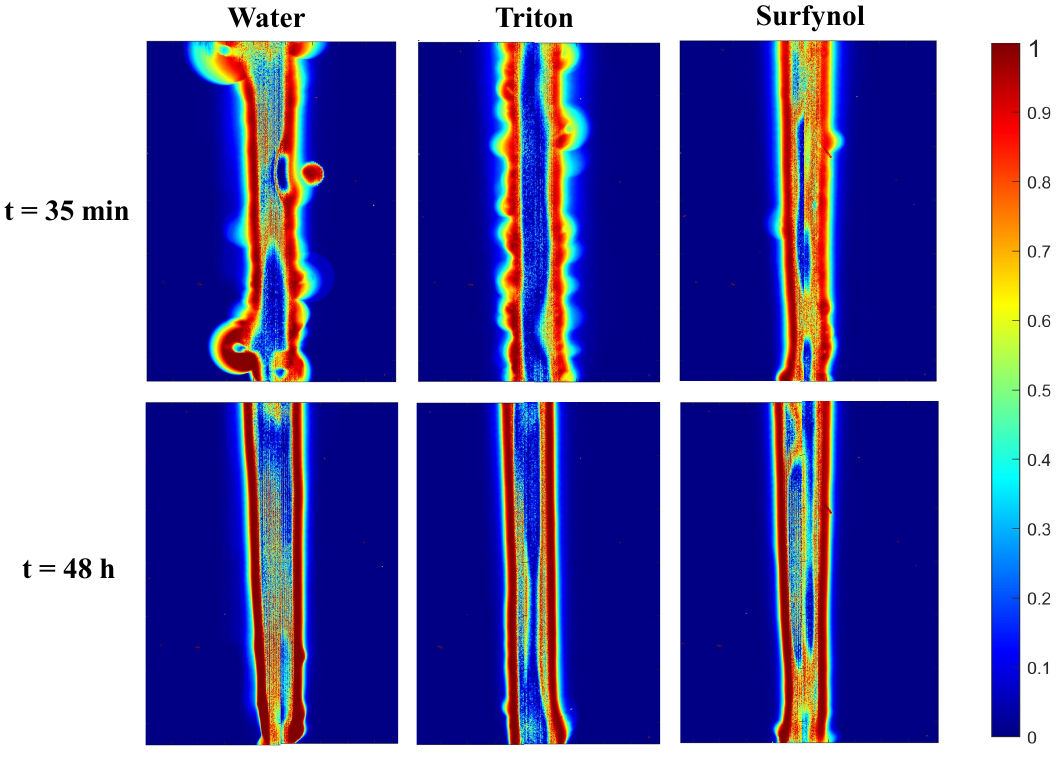}}   
\end{center}
\caption{Time during which a small surface element of the image was occupied by gas. The times were normalized to their corresponding maximum values.}
\label{Heatmap}
\end{figure*}

Figure \ref{Heatmap} shows the (normalized) time during which a small surface element of the image was occupied by gas. At $t=35$ min, the gas occupies a broad, irregular band in the absence of Surfynol 465. With this surfactant, the gaseous band becomes more homogeneous across the entire cathode. At $t=48$ h, the results are similar for the three liquids. 

\begin{figure*}[tbp]
\begin{center}
\resizebox{0.50\textwidth}{!}{\includegraphics{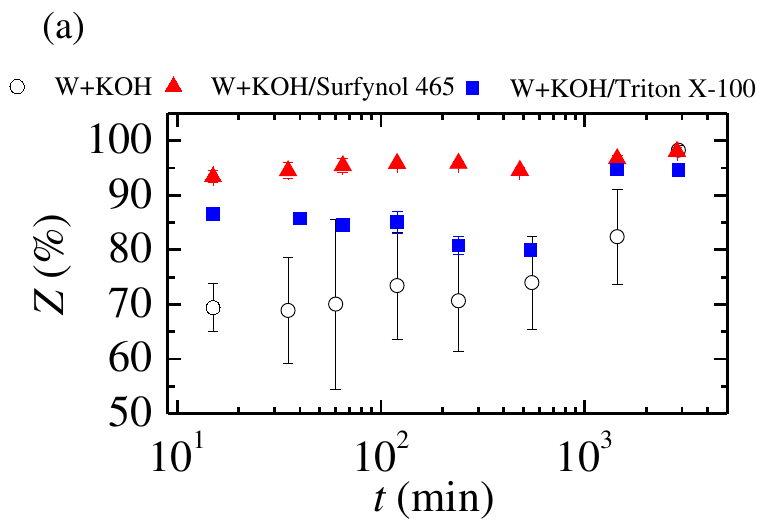}} \resizebox{0.49\textwidth}{!}{\includegraphics{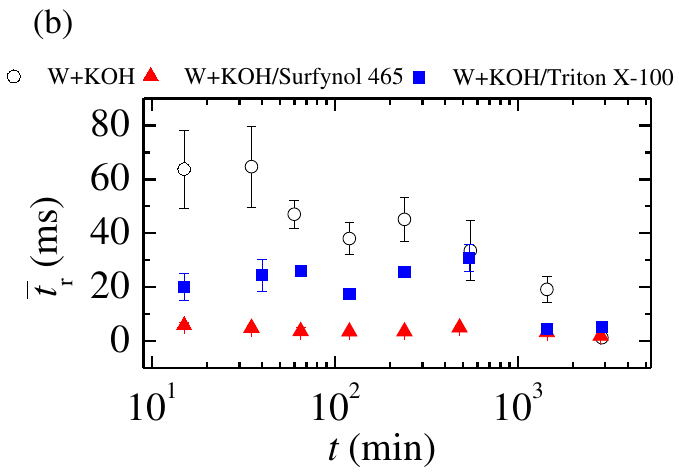}} 
\end{center}
\caption{Temporal evolution of the free-wall percentage $Z$ (a) and the mean residence time $\overline{t}_{r}$ (b) for W+KOH, W+KOH/Triton X-100, and W+KOH/Surfynol 465.}
\label{Zwb}
\end{figure*}

As explained in Sec.\ \ref{sec2}, the free-wall percentage $Z$ indicates the fraction of the electrode surface free of bubbles, while the average residence time $\overline{t}_{r}$ quantifies how long the gas remains trapped on the wall before detaching. Both quantities show that adding a surfactant accelerates the bubble detachment for $t\lesssim 10^3$ min (Fig. \ref{Zwb}). This effect is more noticeable with Surfynol 465. In fact, Surfynol 465 reduced the bubble residence time by one order of magnitude over a significant part of the process.

\textbf{\begin{figure*}[tbp]
\begin{center}
\resizebox{0.50\textwidth}{!}{\includegraphics{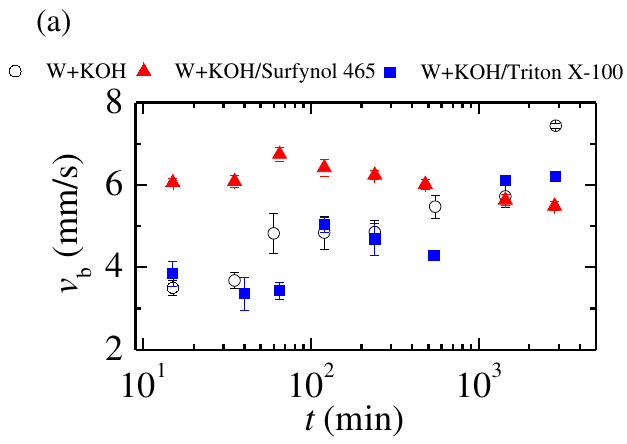}} \resizebox{0.49\textwidth}{!}{\includegraphics{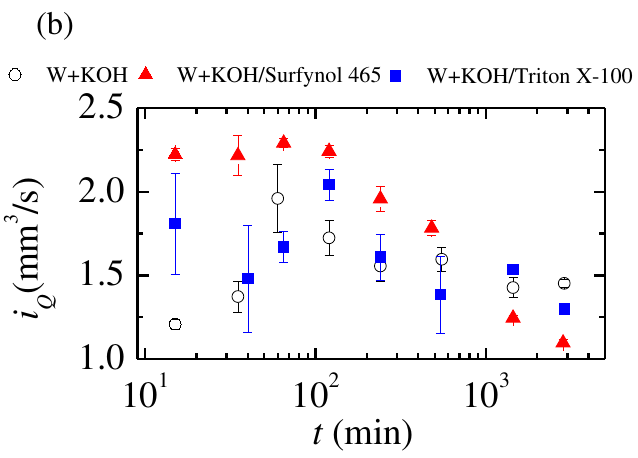}}  
\end{center}
\caption{Temporal evolution of the mean rising bubble velocity $v_b$ (a) and the gas flow rate index $i_{Q}$ (b) for W+KOH, W+KOH/Triton X-100, and W+KOH/Surfynol 465.}
\label{vb}
\end{figure*}}

We also evaluated the average bubble rising velocity $v_b$ from the experimental images (see Sec.\ \ref{sec2}). This allowed us to calculate a measure of the hydrogen production rate ($i_Q$). The bubbles rise faster in the presence of Surfynol 465, resulting in a higher hydrogen production rate. The three liquids behave similarly in the last part of the process.

Finally, we measured the energy performance of the process by calculating the ratio of the hydrogen production rate ($i_Q$) to the consumed electrical power ($I\,V$) (Fig.\ \ref{iEC}). Adding Surfynol 465 improves process performance, except in the late part of the process. 

\textbf{\begin{figure*}[tbp]
\begin{center}
\resizebox{0.5\textwidth}{!}{\includegraphics{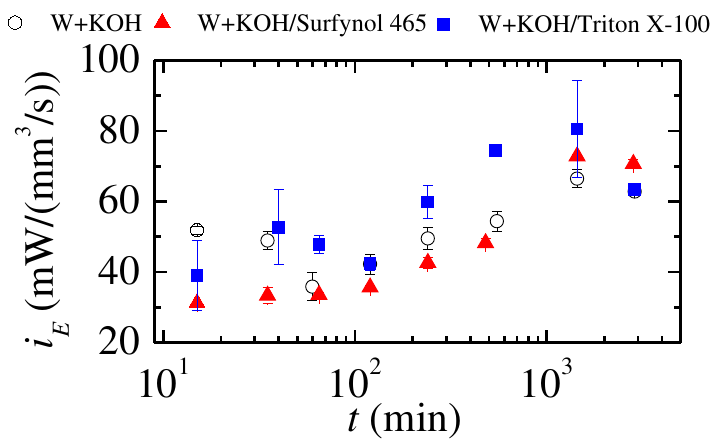}}   
\end{center}
\caption{Temporal evolution of the energy consumption index $i_E$ for W+KOH, W+KOH/Triton X-100, and W+KOH/Surfynol 465.}
\label{iEC}
\end{figure*}}

Overall, the results presented in this section show significant improvements in bubble size distribution, detachment, rising velocity, hydrogen production rate, and energy performance. These improvements are not appreciated after 48 h of use. We hypothesize that this is due to chemical degradation that suppresses the action of Surfynol 465, because there is not recirculation in our simplified electrolyzer model. Section \ref{sec5} will show that when the electrolyte continuously recirculates in an AEMEL electrolyzer, its positive effects persist longer throughout the process.

\subsection{SEM analysis of the electrodes}

We analyzed the surface morphology of the cathodes after 48 hours of use for W+KOH, W+KOH/Triton X-100, and W+KOH/Surfynol 465 using a scanning electron microscope ({\sc Zeiss EVO LS 15}). High-resolution micrographs were acquired to examine the woven structure of the stainless steel mesh/felt and any surface modifications induced by alkaline exposure or surfactant addition.

\begin{figure*}[tbp]
\begin{center}
\resizebox{0.7\textwidth}{!}{\includegraphics{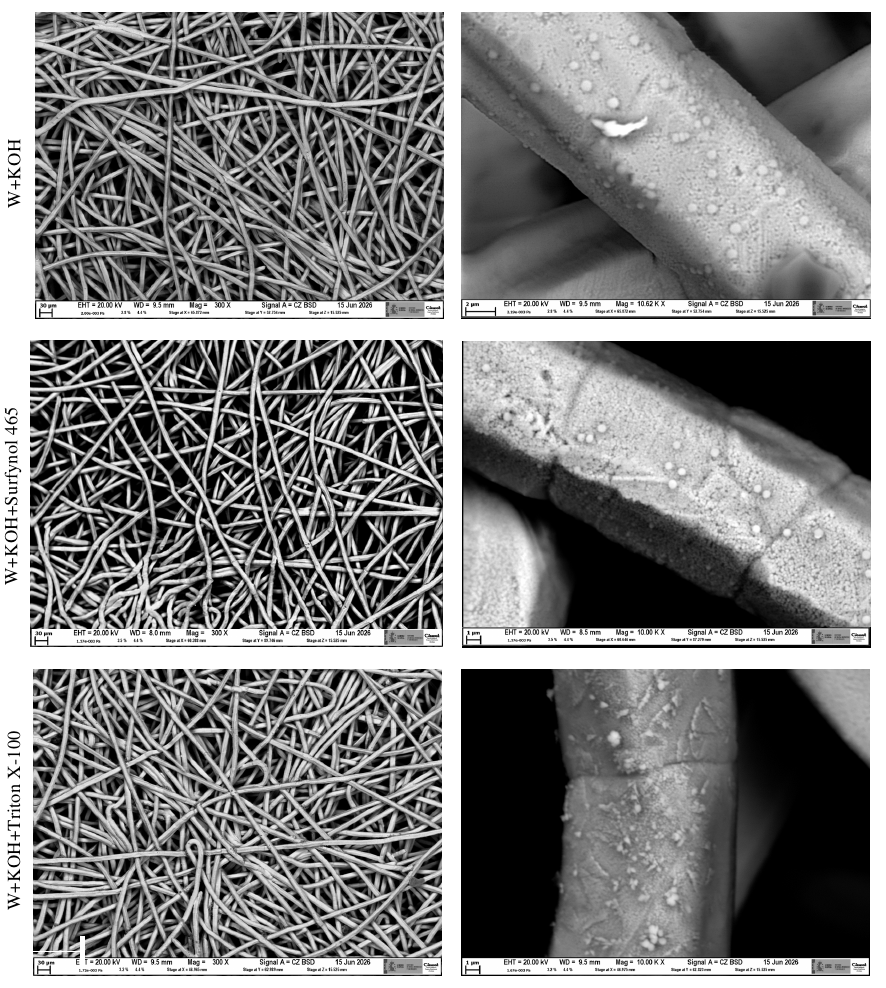}}   
\end{center}
\caption{SEM images of the cathode after 48 hours of use for W+KOH, W+KOH/Triton X-100, and W+KOH/Surfynol 465, after cleaning the precipitates of potassium-rich salts.}
\label{SEM_removingKOH}
\end{figure*}

Figure \ref{SEM_removingKOH} shows no significant changes in the microstructure of the sintered stainless steel fibers that compose the cathode felt. Note that before the SEM, the precipitates of potassium-rich salts were cleaned using an ultrasonic bath in alcohol for 30 min at 60 $^{\circ}$C, followed by air drying.

\begin{figure*}[tbp]
\begin{center}
\resizebox{0.7\textwidth}{!}{\includegraphics{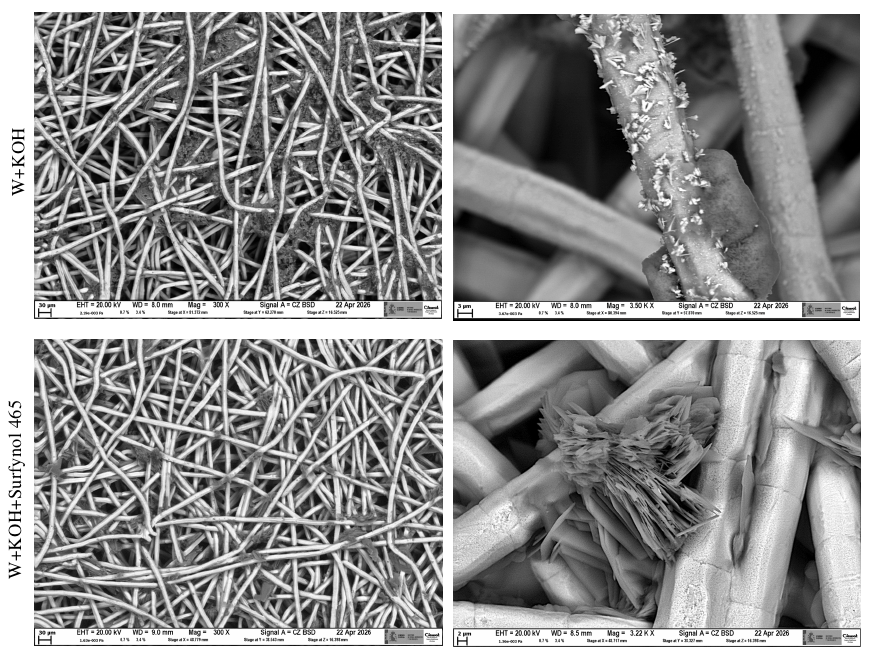}}   
\end{center}
\caption{SEM images of the cathode after 48 hours of use for W+KOH, W+KOH/Triton X-100, and W+KOH/Surfynol 465, before cleaning the precipitates of potassium-rich salts.}
\label{SEM_KOHdeposition}
\end{figure*}

The image without surfactant in Fig.\ \ref{SEM_KOHdeposition} shows a typical stainless steel composition with surface traces of potassium salts, densely spread on the surface. Localized deposits appear along the fiber intersections, consistent with mild alkaline exposure and evaporation during post-operation handling.

With Surfynol 465, the distribution and morphology of the precipitates of potassium-rich salts are totally different, exhibiting a clear lamellar structure. Therefore, the surfactant significantly modifies deposit morphology, producing larger, more irregular surface features compared to the reference electrolyte. The mesh fibers exhibit more heterogeneous coverage, with regions containing thick accumulations of precipitated salts.

These observations support the hypothesis that Surfynol 465 promotes heterogeneous electrolyte distribution on the mesh surface, facilitating the formation of large salt aggregates during operation and subsequent drying.

\section{Results. AEMEL electrolyzer experiments}
\label{sec5}

This section presents the main results of experiments with the AEMEL electrolyzer described in Sec.\ \ref{sec3}, a configuration closer to industrial applications. We used W+KOH and W+KOH/Surfynol 465. We confirmed that conductivity remains within the expected ranges in both cases, indicating that Surfynol 465 does not significantly alter bulk ionic transport.

% Polarization curve
To evaluate the electrochemical performance, we first compare the initial polarization curves with the two electrolytes (Fig.\ \ref{VI_ciemat}). The addition of Surfynol 465 significantly improves cell efficiency across all tested density currents, yielding a consistent voltage reduction of 140-150 mV. Specifically, at 0.1 A/cm$^2$, the cell voltage drops from 1.98 V to 1.83 V, representing a 20\% reduction in overpotential. Similarly, at 0.3 A/cm$^2$ and 0.45 A/cm$^2$, the overpotential decreased around 16\%. At a constant operating voltage of 2 V, the surfactant increases the current density nearly threefold, from 0.13 A/cm$^2$ to 0.37 A/cm$^2$.

\textbf{\begin{figure*}[tbp]
\begin{center}
\resizebox{0.5\textwidth}{!}{\includegraphics{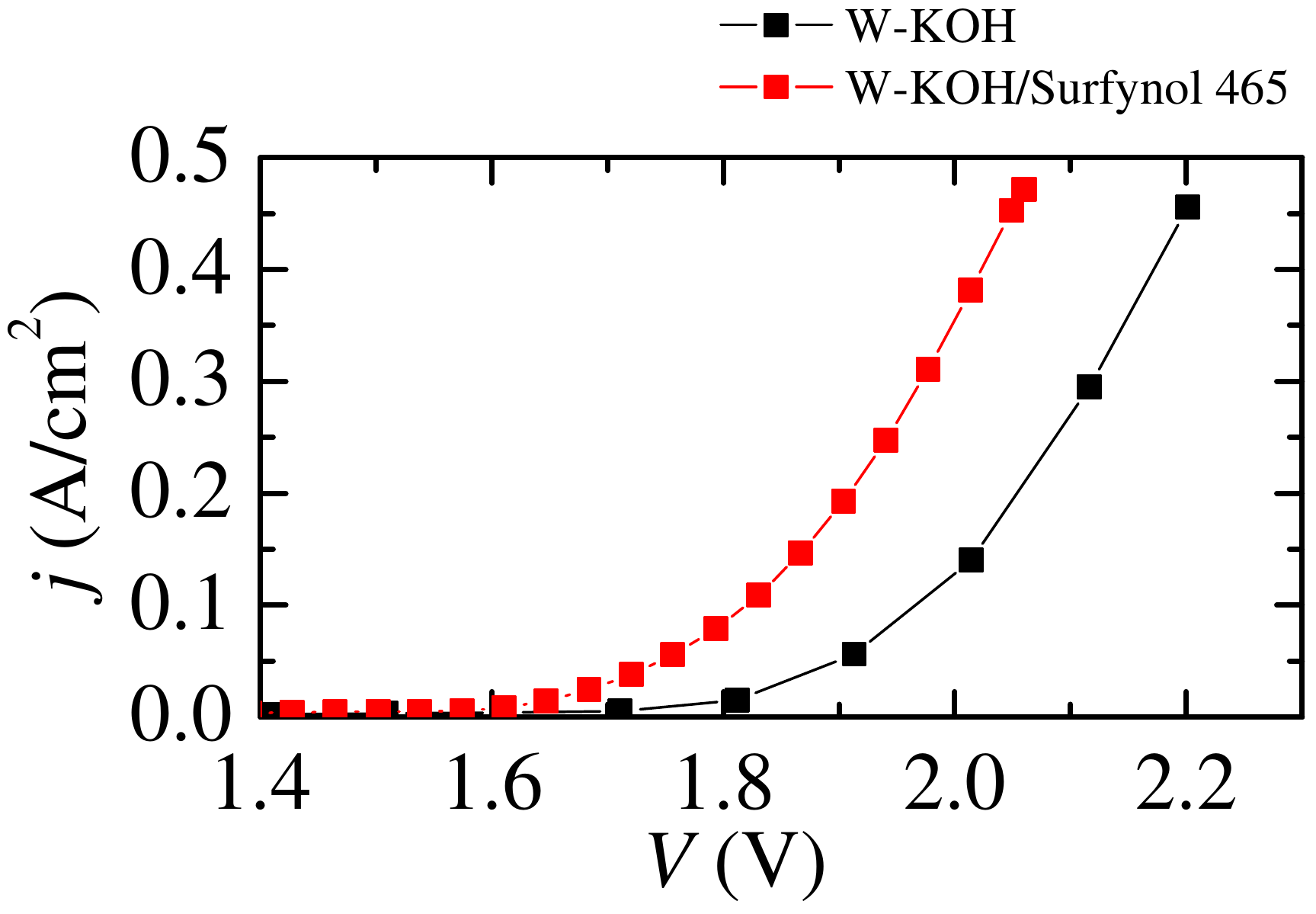}}   
\end{center}
\caption{Initial polarization curves for W+KOH and W+KOH/Surfynol 465.}
\label{VI_ciemat}
\end{figure*}}

% Cell power
Now, we focus on the cell performance for the 7-day operational test. Figure \ref{P_ciemat} shows the instantaneous cell power $P$ for W+KOH and W+KOH/Surfynol 465. For W+KOH, the power source maintained the voltage control after the initial activation period. Conversely, in the case of  W+KOH/Surfynol 465, the power supply reached its current delivery limit. When this occurs, the system no longer regulates the voltage, which drifts to fix the current at its maximum setpoint. We verified that the power performance was effectively $100\%$, i.e., the power consumed was equal to the power generated. The mean cell power $P$ over one week increased by around $40\%$ when using Surfynol 465. This extraordinary improvement is consistent with the results of high-speed optical diagnostics (Sec.\ \ref{sec4}). Please, note that the experiments were developed at room temperature, and the temperature can slightly influence our results. 

\textbf{\begin{figure*}[tbp]
\begin{center}
\resizebox{0.5\textwidth}{!}{\includegraphics{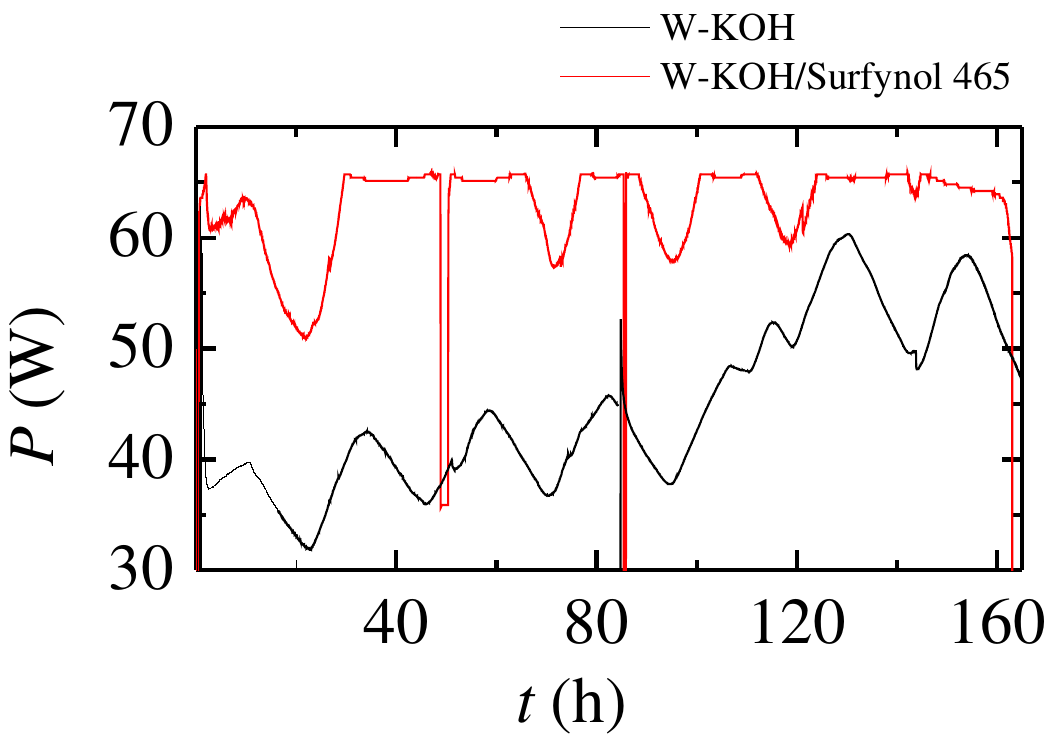}}   
\end{center}
\caption{Instantaneous cell power $P$ obtained for W+KOH and W+KOH/Surfynol 465.}
\label{P_ciemat}
\end{figure*}}

% Corrosion kinetics
Material degradation under alkaline conditions was quantified using downstream electrochemical corrosion probes (Fig.\ \ref{corrosion}). Overall, these results highlight the importance of material selection when considering surfactant-assisted operation. Surfynol 465 reduces corrosion in brass (Cu-based alloy) and stainless steel (inox), but accelerates degradation in carbon steel (Fe-based alloy). Thus, the corrosion analysis reveals a strong alloy-dependent response to Surfynol 465. Muntz metal (a Cu-based alloy) exhibits reduced corrosion and a transition from parabolic to linear kinetics, consistent with the partial formation of an organic film. Stainless steel (Inox) shows a dramatic reduction in corrosion rate, suggesting enhanced passive film stability or reduced interfacial reactivity. In contrast, carbon steel (Fe-based alloys) experiences accelerated corrosion, indicating that Surfynol 465 disrupts its natural alkaline passivation. This alloy-dependent behavior underscores the need to evaluate long-term material compatibility in AEMEL systems when surfactants are introduced into the electrolyte. This is beyond the scope of this work.

\textbf{\begin{figure*}[tbp]
\begin{center}
\resizebox{0.5\textwidth}{!}{\includegraphics{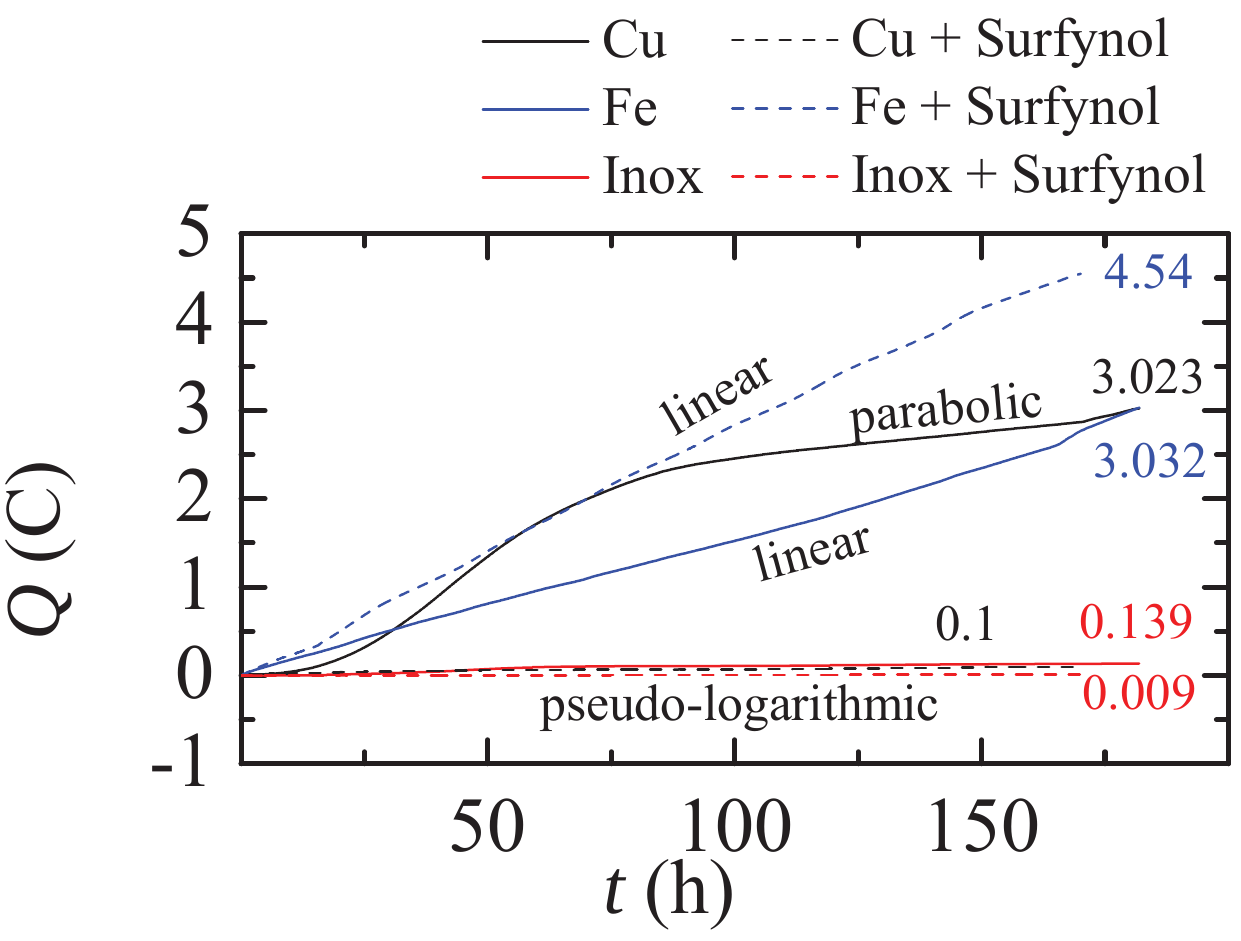}}   
\end{center}
\caption{Corrosion charge $Q$ for Cu-based alloy, Fe-based alloy, and stainless steel (Inox).}
\label{corrosion}
\end{figure*}}

Finally, we assess gas purity and volatile contaminants. Gas chromatography of the hydrogen produced during the W+KOH/Surfynol 465 experiment confirmed high purity (around $100\%$), with only trace CO$_2$ detected and no significant contaminants. The absence of volatile organic species in the gas stream suggests that the possible Surfynol 465 decomposition products remain confined to the liquid phase and do not contaminate the hydrogen. This observation is consistent with the SEM results.

\section{Conclusions}
\label{sec6}
This work demonstrates the crucial role of surfactant adsorption kinetics in mitigating bubble-induced efficiency losses in water electrolysis. Using Surfynol 465 as an ultra-fast kinetic surfactant candidate, we establish a direct link between dynamic surface tension and macroscopic electrolyzer performance. High-speed optical diagnostics in an AWE laboratory-scale cell reveal that Surfynol 465 reduces bubble residence time on the cathode by an order of magnitude. It effectively prevents the growth of large hydrogen bubbles ($>200\ \mu$m) compared to the slower surfactant Triton X-100 and the surfactant-free baseline electrolyte. This rapid detachment increases the mean bubble-rise velocity, thereby significantly enhancing the hydrogen production rate during the early stages of operation.

We validated the previous results on a larger-scale AEMEL test bench. These microscopic improvements translate into a significant performance boost. Adding Surfynol 465 to the 1 M KOH electrolyte decreases the cell overpotential by 140--150 mV. At a constant operating voltage of 2 V, the surfactant nearly triples the current density from 0.13 to 0.37 A/cm$^2$. Furthermore, the beneficial effects of the surfactant persist during continuous operation with recirculating electrolyte, resulting in a 40\% increase in average cell power during a week-long test. 

Material degradation under alkaline conditions was quantified using downstream corrosion analysis. The resulting measurements reveal a strong alloy-dependent response to Surfynol 465. The surfactant reduces corrosion kinetics in stainless steel and brass, likely by forming a protective organic film. In contrast, it accelerates degradation in carbon steel by disrupting its natural alkaline passivation. Finally, gas chromatography also confirms that the hydrogen produced remains highly pure and free of volatile contaminants.

Overall, utilizing surfactants with ultrafast adsorption kinetics represents a highly effective, low-cost passive strategy to boost electrolyzer efficiency, provided that stack materials are carefully selected to ensure long-term chemical compatibility.

\vspace{1cm}

 Support from the Spanish Ministry of Science and Innovation (MCIN) through grant PID2025-171659NB-I00/ AEI// FEDER, UE is gratefully acknowledged. This work was co-funded (85\%) by the European Union through the European Regional Development Fund (FEDER) and the Regional Government of Extremadura. Managing Authority: Ministry of Finance (Grant GR24077).

\vspace{1cm}

%\bibliographystyle{elsarticle-num-names} 
%\bibliography{central,central2}

\end{document}